\documentclass[conference]{IEEEtran}
\usepackage[draft]{hyperref}
\usepackage{orcidlink}
\IEEEoverridecommandlockouts
\usepackage{cite}
\usepackage{url}
\usepackage{amsmath,amssymb,amsfonts}
\usepackage{algorithmic}
\usepackage{graphicx}
\usepackage{textcomp}
\usepackage{xcolor}
\usepackage{booktabs}
\usepackage{multirow}
\usepackage{tikz}
\usepackage{pgfplots}
\pgfplotsset{compat=1.17}
\usetikzlibrary{arrows.meta, positioning, shapes.geometric, fit, backgrounds, calc, shadows}
\def\BibTeX{{\rm B\kern-.05em{\sc i\kern-.025em b}\kern-.08em
    T\kern-.1667em\lower.7ex\hbox{E}\kern-.125emX}}
\begin{document}
\bstctlcite{IEEEexample:BSTcontrol}

\title{Speaker-Aware Temporal Aggregation Strategies on Segment Representations for Depression Detection in Dyadic Interaction: A Benchmark Study}

\author{
\small
\IEEEauthorblockN{
Anisha Pattanayak\orcidlink{0009-0005-2556-4472}$^{\star}$,
Huang-Cheng Chou\orcidlink{0000-0003-2125-5689}$^{\star\dagger}$,
Shrikanth Narayanan\orcidlink{0000-0002-1052-6204}$^{\star\dagger}$,
Sudarsana Reddy Kadiri\orcidlink{0000-0001-5806-3053}$^{\star\dagger}$
}
\IEEEauthorblockA{
$^{\star}$\textit{Ming Hsieh Department of Electrical and Computer Engineering, University of Southern California, USA}\\
$^{\dagger}$\textit{Signal Analysis and Interpretation Laboratory (SAIL), University of Southern California, USA}
}
}
\maketitle
\begin{abstract}
Speech-based depression detection compresses features from short audio segments into one speaker-level decision, a step called temporal aggregation rarely studied on its own.
Most benchmarks fix a single self-supervised encoder and a single hand-picked layer, so a reported gain may reflect the pipeline rather than the aggregation method itself.
We introduce \textbf{DEPOOL}, a controlled benchmark that compares six aggregation architectures with six frozen speech backbones on an English and a Mandarin depression corpus, where each configuration learns which backbone layers matter rather than fixing one by hand.
Across the resulting 72-configuration grid, a third of configurations collapse into predicting a single class for every speaker, a failure tied to the backbone as much as to the method, and the architecture that is most stable in a single-seed run becomes unreliable when training repeats across seeds.
Robustness to backbone and seed, rather than average accuracy across a single pipeline, should be a first-class benchmarking criterion for temporal aggregation in clinical speech.
\end{abstract}


\section{Introduction}
Automatic depression detection from speech is attracting growing interest as a complement to clinical screening. Depression affects hundreds of millions of people worldwide, and substantial treatment gaps remain~\cite{who2023depression}; speech is also naturally available during many clinical interactions and has therefore been studied as a potential source of behavioral markers~\cite{low2020automated}.

Most systems in this space share a common structure in which a feature extractor processes fixed-length audio segments and a classifier turns those segment features into a single speaker-level prediction for the interview. This last step, \emph{aggregation}, contains important open design choices. Depression-related acoustic characteristics can vary over the course of an interaction~\cite{cummins2015review}, so an aggregation method that models variation across segments may behave differently from one that averages all segments uniformly.

A subtler problem sits underneath this one. Many existing comparisons of aggregation methods use one self-supervised learning (SSL) backbone (a large speech model pre-trained on unlabeled audio) and one transformer layer selected in advance or through preliminary probing. This makes it difficult to determine whether an observed advantage belongs to the aggregation architecture or to the particular backbone and layer with which it was paired. This paper asks whether the relative ranking of temporal aggregation strategies for depression detection persists when the SSL backbone, layer combination, and corpus vary.

To answer this, we built \textbf{DEPOOL} and trained all six aggregation architectures on all six SSL backbones on both E-DAIC~\cite{gratch2014distress} and MODMA~\cite{cai2020modma}, forming a 72-cell grid. To remove hand-picked layer selection as a confound, every configuration learns a softmax-weighted combination of all hidden layers in its backbone rather than using a single fixed layer. All splits are speaker-independent and use stratified, participant-grouped folds so that no segment from a test-set speaker ever appears in training. The protocol is informed by the layer-featurization and cross-backbone evaluation practices of the SUPERB benchmark family~\cite{yang2021superb,wu2024openemotion}, providing a consistent and reproducible evaluation design across model choices.
This paper provides several contributions.
\begin{itemize}
    \item We present a controlled cross-product benchmark of aggregation architecture, SSL backbone, and corpus rather than a single-pipeline comparison.
    \item We introduce a semi-fine-tuned layer-aggregation protocol that removes hand-picked-layer selection as a confound.
    \item We report the finding that roughly a third of the 72 configurations collapse into single-class prediction, a phenomenon invisible to any study reporting one backbone, and we show that this collapse is seed-sensitive as well as backbone-dependent.
    \item We provide strictly speaker-independent, stratified 60/20/20 splits and an open release of the pipeline, splits, and results\footnote{\protect\url{https://github.com/anishxa/Temporal_Pooling_Benchmark}}.
\end{itemize}

\begin{figure}[!t]
\centering
\begin{tikzpicture}[
  stage/.style={rectangle, rounded corners=4pt, draw=black!55, line width=0.6pt,
                text width=1.7cm, align=center, font=\scriptsize,
                minimum height=1.0cm, minimum width=1.7cm,
                fill=white, drop shadow},
  audioblk/.style={stage, fill=blue!8},
  frozenblk/.style={stage, fill=orange!12},
  learnblk/.style={stage, fill=green!10},
  outblk/.style={stage, fill=red!10},
  arr/.style={-{Stealth[length=3pt]}, thick, black!55},
  lbl/.style={font=\tiny, black!55, align=center},
  stagelbl/.style={font=\scriptsize\bfseries, black!45, fill=white, inner sep=1pt, align=center}
]
\node[audioblk] (audio) at (0,0) {Audio\\(E-DAIC/\\MODMA)};
\node[audioblk, right=0.3cm of audio] (seg) {3s clips\\1.5s stride};
\node[frozenblk, right=0.3cm of seg] (SSL) {Frozen SSL\\(1 of 6)};

\draw[arr] (audio) -- (seg);
\draw[arr] (seg) -- (SSL);

\node[frozenblk] (feat) at (0,-2.2) {Hidden\\states\\$[L,D]$};
\node[learnblk, right=0.3cm of feat] (fz) {Softmax\\weights};
\node[learnblk, right=0.3cm of fz] (agg) {Agg.\\head};

\draw[arr] (feat) -- (fz);
\draw[arr] (fz) -- (agg);
\draw[arr] (SSL.south) -- ++(0,-0.4) -| (feat.north);

\begin{scope}[on background layer]
\node[fit=(fz)(agg), inner sep=3pt, rounded corners=6pt, fill=green!4, draw=green!25, line width=0.5pt] {};
\end{scope}

\node[outblk] (out) at (1.15,-4.4) {Speaker\\prediction};
\draw[arr] (agg.south) -- ++(0,-0.4) -| (out.north);

\node[stagelbl] at ($(seg.north)+(0,0.6)$) {STEP 1: Segment \& encode};
\node[stagelbl] at ($(fz.north)+(0,0.6)$) {STEP 2: Learn \& pool};
\node[stagelbl] at ($(out.north)+(0,0.4)$) {STEP 3: Predict};

\node[lbl] at (1.15,-5.2) {Repeated for all 72 combinations};
\end{tikzpicture}
\vspace{-3mm}
\caption{DEPOOL pipeline: Components are color-coded by training status.}
\label{fig:pipeline}
\vspace{-1mm}
\end{figure}
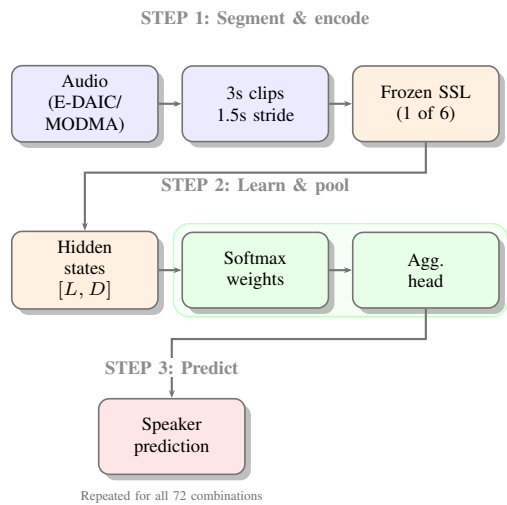

\section{Related Work}
\subsection{Depression detection from speech}
Early systems relied on hand-crafted acoustic features, including Mel-frequency cepstral coefficients (MFCCs), pitch, jitter, and shimmer, paired with support vector machines or logistic regression~\cite{cummins2015review,albuquerque2021association}; bidirectional long short-term memory (LSTM) networks later showed that modeling temporal dynamics can improve over static summaries~\cite{alhanai2018detecting}. Self-supervised learning (SSL) models such as HuBERT (Hidden-unit Bidirectional Encoder Representations from Transformers, or Hidden-unit BERT)~\cite{hsu2021hubert}, WavLM~\cite{chen2022wavlm}, data2vec~\cite{baevski2022data2vec}, and XLS-R (a large-scale cross-lingual speech representation model)~\cite{babu2021xlsr} learn transferable representations from unlabeled audio. Such representations have been studied for depression detection~\cite{zhang2021depa} and related paralinguistic tasks such as speech emotion recognition~\cite{pepino2021emotion}.
The SUPERB (Speech processing Universal PERformance Benchmark)~\cite{yang2021superb} showed that an SSL backbone's apparent quality depends heavily on the task and probing protocol used to evaluate it, and that no single backbone dominates across the board; EMO-SUPERB made a similar reproducibility argument for speech emotion recognition, releasing a unified, leakage-safe benchmark codebase across backbones~\cite{wu2024openemotion}. Layer-wise studies further show intermediate transformer layers often encode prosodic and affective information better than layers near the input or output~\cite{maji2024investigation,han2024croSSLingual}, motivating our choice to learn layer weights rather than fix them by hand.

\subsection{Segment aggregation in clinical speech}
The aggregation step itself has received comparatively little attention, usually confined to a single backbone. Common approaches include majority voting over segment predictions, mean pooling of segment embeddings, or a single recurrent model over the segment sequence~\cite{alhanai2018detecting,ma2016depaudionet}.
Ringeval et al.~\cite{ringeval2019avec} fused frame-level predictions with simple statistics; Ma et al.~\cite{ma2016depaudionet} proposed DepAudioNet, a convolutional neural network (CNN) followed by an LSTM, with mean pooling as its aggregation step. Attention-based aggregation has been studied in related paralinguistic tasks, including local attention over LSTM outputs for emotion recognition~\cite{mirsamadi2017automatic}, which relates to the self-attention and GRU-with-attention heads studied here. NetVLAD (Vector of Locally Aggregated Descriptors)~\cite{arandjelovic2016netvlad} and attentive statistics pooling~\cite{okabe2018attentive} are standard for speaker embeddings; to our knowledge, their robustness for depression screening has not been compared systematically across multiple SSL backbones.
A learnable classification (CLS) token, originating in BERT-style text classification~\cite{devlin2019bert}, also provides a general mechanism for pooling a variable-length sequence and is considered here.

\section{The DEPOOL Benchmark}
\subsection{Datasets}
We use two depression corpora that differ in language, elicitation protocol, and population, so a finding that holds on both is less likely to be an artifact of one dataset, recording setup, or annotation style.

\textbf{E-DAIC}~\cite{ringeval2019avec,gratch2014distress} contains semi-structured English clinical interviews with a virtual agent, labeled by the 8-item Patient Health Questionnaire (PHQ-8), a self-report screening scale binarized at PHQ-8~$\geq$~10 (screen-positive). After speaker-independent splitting (Section~\ref{sec:splitting}), our partition holds 122 participants: 75 for training, 24 for validation, and 23 held out (17 healthy, 6 depressed) for final evaluation. This is the subset with complete audio and labels after preprocessing, re-split rather than using the official challenge partition, so absolute numbers are not comparable to challenge leaderboards.

\textbf{MODMA}~\cite{cai2020modma} is a Mandarin corpus of clinically diagnosed major depressive disorder (MDD) patients and matched healthy controls (HC), recorded during structured interview and reading tasks, with clinician rather than self-report diagnoses. The same splitting procedure yields 52 participants: 31 training, 11 validation, and 10 held out (5 healthy, 5 depressed), with the same three-way roles as E-DAIC.

\subsection{Segmentation and Preprocessing}
Participant-only speech is isolated from each interview; interviewer or virtual-agent turns are excluded so that predictions are based on the target participant's speech. Audio is resampled to 16~kHz mono to standardize the sampling rate across corpora. We slide a 3-second window across each participant turn with a 1.5-second stride (50\% overlap). This fixed segment length captures short stretches of connected speech without requiring a backbone pass over the full interview, while overlap reduces sensitivity to arbitrary window boundaries. Turns shorter than one window are dropped by the preprocessing pipeline. Each clip is centered and scaled using its own mean and standard deviation, which normalizes amplitude offset and scale but does not eliminate all channel or corpus differences, and is cropped or zero-padded to exactly 48{,}000 samples (3~s at 16~kHz) before backbone inference. An interview is represented downstream as an ordered sequence of 3-second clips.

\subsection{SSL Backbones}
DEPOOL evaluates six frozen speech backbones (Appendix Table~\ref{tab:backbones}): WavLM-Base-Plus, WavLM-Large, HuBERT-Large, Wav2Vec2-Robust, Data2Vec-Audio-Large, and XLS-R-1B. These checkpoints span multiple model sizes, pretraining objectives, and training corpora. ``Frozen'' means that backbone weights are never updated for this task; only the layer-aggregation module (Section~\ref{sec:featurizer}) and downstream aggregation head (Section~\ref{sec:aggregation}) are trained. HuBERT-Large and Data2Vec-Audio-Large are ASR-fine-tuned checkpoints rather than purely self-supervised checkpoints, so observed backbone differences may reflect ASR fine-tuning as well as architecture, scale, pretraining objective, and pretraining data.

For every clip, we extract the full stack of hidden states and mean-pool each layer over the backbone's frame axis, giving an $[L,D]$ representation per clip, where $L$ and $D$ are listed for each checkpoint in Appendix Table~\ref{tab:backbones}.

\subsection{Semi-Fine-Tuned Layer Aggregation}
\label{sec:featurizer}
Different backbone layers can encode different information, and selecting one layer in advance can bias a comparison. Instead, each independently trained configuration learns a soft weighted average of all $L$ layers, following the featurizer design used in SUPERB-style benchmarks~\cite{yang2021superb}. For clip $t$, with per-layer vectors $\{\mathbf{h}^{(1)}_t,\ldots,\mathbf{h}^{(L)}_t\}$, $\mathbf{h}^{(\ell)}_t\in\mathbb{R}^{D}$:
\begin{equation}
\tilde{\mathbf{h}}_t = \sum_{\ell=1}^{L} \text{softmax}(w_\ell)\, \mathbf{h}^{(\ell)}_t, \qquad \mathbf{z}_t = \mathbf{W}_p \tilde{\mathbf{h}}_t + \mathbf{b}_p,
\label{eq:featurizer}
\end{equation}
where the symbols are defined as follows:
\begin{itemize}
\item $t \in \{1,\ldots,T\}$ indexes the 3-second clip within an interview;
\item $\ell \in \{1,\ldots,L\}$ indexes the hidden-state layer, with $L$ listed for each checkpoint in Appendix Table~\ref{tab:backbones};
\item $\mathbf{h}^{(\ell)}_t \in \mathbb{R}^{D}$ is the frame-mean-pooled hidden state of layer $\ell$ for clip $t$, with $D$ listed for each checkpoint in Appendix Table~\ref{tab:backbones};
\item $w_\ell \in \mathbb{R}$ is one learnable scalar per layer, shared across clips and interviews within one trained configuration;
\item $\text{softmax}(w_\ell) = \exp(w_\ell)/\sum_{\ell'=1}^{L}\exp(w_{\ell'})$ normalizes the $L$ layer weights to be non-negative and sum to one;
\item $\tilde{\mathbf{h}}_t \in \mathbb{R}^{D}$ is the resulting layer-weighted combination of hidden states for clip $t$;
\item $\mathbf{W}_p \in \mathbb{R}^{256\times D}$ is a learnable projection matrix for the current configuration;
\item $\mathbf{b}_p \in \mathbb{R}^{256}$ is the corresponding learnable bias term;
\item $\mathbf{z}_t \in \mathbb{R}^{256}$ is the resulting clip embedding passed to the aggregation architectures of Section~\ref{sec:aggregation}.
\end{itemize}
The softmax constrains the coefficients to be non-negative and to sum to one. We use this normalized form to preserve a comparable feature scale while learning the relative contribution of each layer, following the SUPERB featurizer design~\cite{yang2021superb}.
For each backbone--architecture configuration, $\{w_\ell\}$, $\mathbf{W}_p$, $\mathbf{b}_p$, and the downstream head are trained jointly; these parameters are not shared across independently trained configurations. Every head receives a 256-dimensional clip embedding regardless of the backbone, removing hidden-size differences at the head input while leaving other backbone differences intact.

\subsection{Aggregation Architectures}
\label{sec:aggregation}
Each head takes the ordered clip embeddings $\{\mathbf{z}_1,\ldots,\mathbf{z}_T\}$, $\mathbf{z}_t\in\mathbb{R}^{256}$, of one interview and a validity mask $\mathbf{m}\in\{0,1\}^T$ ($m_t=1$ for a real clip, 0 for padding), and compresses them into one speaker vector that a classifier maps to depressed vs.\ healthy. Let $\mathcal{S}=\{t: m_t=1\}$.
\textit{Mean Pooling.} $\mathbf{p}_{\text{mean}} = \frac{1}{|\mathcal{S}|}\sum_{t\in\mathcal{S}}\mathbf{z}_t$.
\textit{Statistical Pooling.} Element-wise mean, standard deviation, maximum, and minimum over $\{\mathbf{z}_t\}_{t\in\mathcal{S}}$, concatenated: $\mathbf{p}_{\text{stat}} = [\mu;\sigma;\max;\min]\in\mathbb{R}^{1024}$.
\textit{Self-Attention Pooling.} Learns a per-clip weight $\alpha_t$:
\begin{equation}
\alpha_t = \frac{\exp(\mathbf{v}^\top\tanh(\mathbf{W}\mathbf{z}_t+\mathbf{b})) m_t}{\sum_{t'}\exp(\mathbf{v}^\top\tanh(\mathbf{W}\mathbf{z}_{t'}+\mathbf{b})) m_{t'}}, \ \mathbf{p}_{\text{attn}}=\sum_{t\in\mathcal{S}}\alpha_t \mathbf{z}_t.
\label{eq:attn}
\end{equation}
$\mathbf{W}\in\mathbb{R}^{128\times256}$ and $\mathbf{b}\in\mathbb{R}^{128}$ map each clip into a 128-dimensional attention space, $\mathbf{v}\in\mathbb{R}^{128}$ scores it, and $m_t$ masks padding before normalization, giving $\mathbf{p}_{\text{attn}}\in\mathbb{R}^{256}$.
\textit{Transformer Encoder.} $\mathbf{z}_t$ is projected to a 64-dimensional bottleneck with a rectified linear unit (ReLU) and dropout~0.3. A single-layer, 4-head encoder (dropout~0.3) then processes a prepended learnable classification (CLS) token and the clip embeddings, with padding excluded by a key-padding mask. No positional encodings are added, so the pooled CLS representation is permutation-invariant over valid clips.
\textit{Bidirectional GRU with Attention.} A two-layer bidirectional gated recurrent unit (GRU; hidden size 128 per direction, 256-dim concatenated output) scans the sequence in both directions, followed by Eq.~\eqref{eq:attn} applied to the GRU outputs in place of $\mathbf{z}_t$.
\textit{NetVLAD.} Each $\mathbf{z}_t$ is projected to $\mathbf{q}_t\in\mathbb{R}^{64}$; a learned linear assignment with softmax gives $\bar{a}_k(\mathbf{q}_t)$ over $K=2$ clusters. Residuals $\sum_{t\in\mathcal{S}}\bar{a}_k(\mathbf{q}_t)(\mathbf{q}_t-\boldsymbol{\mu}_k)$ are intra-normalized, concatenated, and L2-normalized, following~\cite{arandjelovic2016netvlad}.
Every architecture ends in a small multi-layer perceptron (MLP) classifier with dropout and one ReLU hidden layer emitting two class logits; softmax probabilities are used only for evaluation.
Excluding the configuration-specific layer featurizer, head plus classifier parameter counts are approximately: Mean 16.6K, Statistical 131.5K, Self-Attention 49.6K, Transformer 299.8K, Bi-GRU 626.0K, NetVLAD 20.9K. Capacity alone does not predict collapse: Bi-GRU is the largest head yet has no collapsed runs in the single-seed grid (Section~\ref{sec:collapse}).

\subsection{Speaker-Independent Splitting}
\label{sec:splitting}
SSL representations encode substantial speaker-identity information alongside paralinguistic content, so leaking clips from one speaker across the training, validation, and test subsets could let a classifier learn to recognize individual voices rather than depression markers, silently inflating reported performance. We therefore use \texttt{StratifiedGroupKFold}~\cite{pedregosa2011scikit}, a splitting procedure that groups all instances belonging to the same entity (here, participant ID) so they cannot be separated across folds, while additionally stratifying by depression label so each split keeps a similar depressed/healthy ratio. This produces a 60/20/20 train/validation/test partition in which every clip from a given participant falls entirely into one split, matching recommended practice for clinical speech datasets~\cite{schuller2021interspeech}. The training split is used to fit model weights, the validation split only to select the best training checkpoint in the focal runs (Section~\ref{sec:training}), and the test split is held out and evaluated exactly once per configuration.

\begin{table}[!t]
\caption{Mean Metrics by Architecture, Averaged over 6 Backbones ($n=6$ per row; test speakers per corpus: E-DAIC $n=23$, MODMA $n=10$)}
\label{tab:results}
\vspace{-4mm}
\begin{center}
\scriptsize
\setlength{\tabcolsep}{3pt}
\begin{tabular}{llcccccc}
\toprule
\textbf{Set} & \textbf{Architecture} & \textbf{Acc} & \textbf{F1} & \textbf{AUC} & \textbf{Sens} & \textbf{Spec} & \textbf{$U$} \\
\midrule
\multirow{6}{*}{\rotatebox{90}{E-DAIC}}
 & Mean Pooling       & \textbf{0.630} & \textbf{0.534}$\pm$0.073 & 0.613 & 0.750 & 0.588 & 0.696 \\
 & Statistical Pool.  & 0.572 & 0.511$\pm$0.058 & 0.655 & 0.806 & 0.490 & \textbf{0.700} \\
 & Self-Attention     & 0.551 & 0.507$\pm$0.086 & 0.598 & 0.778 & 0.471 & 0.675 \\
 & Bi-GRU + Attn      & 0.601 & 0.508$\pm$0.079 & \textbf{0.662} & 0.722 & \textbf{0.559} & 0.668 \\
 & NetVLAD            & 0.413 & 0.470$\pm$0.093 & 0.607 & \textbf{0.889} & 0.245 & 0.674 \\
 & Transformer Enc.   & 0.529 & 0.290$\pm$0.206 & 0.577 & 0.556 & 0.520 & 0.544 \\
\midrule
\multirow{6}{*}{\rotatebox{90}{MODMA}}
 & Mean Pooling       & 0.717 & 0.745$\pm$0.074 & 0.647 & 0.767 & 0.667 & 0.733 \\
 & Statistical Pool.  & 0.650 & 0.708$\pm$0.042 & 0.537 & 0.800 & 0.500 & 0.700 \\
 & Self-Attention     & 0.717 & 0.731$\pm$0.049 & 0.680 & 0.733 & 0.700 & 0.722 \\
 & Bi-GRU + Attn      & \textbf{0.800} & \textbf{0.811}$\pm$0.093 & \textbf{0.713} & 0.833 & \textbf{0.767} & \textbf{0.811} \\
 & NetVLAD            & 0.733 & 0.632$\pm$0.302 & 0.677 & 0.567 & \textbf{0.900} & 0.678 \\
 & Transformer Enc.   & 0.517 & 0.667$\pm$0.000 & 0.433 & \textbf{0.967} & 0.067 & 0.667 \\
\bottomrule
\end{tabular}
\end{center}
{\footnotesize Acc = accuracy; F1 = positive-class F1; AUC = ROC-AUC; Sens/Spec = sensitivity/specificity as defined in Section~\ref{sec:training}; $U$ = the utility score of Eq.~\eqref{eq:utility}. Each row is the mean over the 6 backbones summarized in Appendix Table~\ref{tab:backbones}, with one run per backbone; F1 is mean$\pm$standard deviation across those runs. Bold marks the best value in each column within a corpus block.}
\end{table}

\subsection{Training and Evaluation}
\label{sec:training}
All 72 configurations are trained independently for 40 epochs with AdamW (learning rate $10^{-3}$, weight decay $10^{-4}$), class-weighted cross-entropy, and random seed 42. The checkpoint with the highest validation positive-class F1 is retained, and the held-out test split is evaluated once using that checkpoint (focal runs; see Section~\ref{sec:limitations} for the single-seed grid). We report accuracy, positive-class F1, area under the receiver-operating-characteristic curve (ROC-AUC), sensitivity (recall of the screen-positive/MDD class), and specificity (recall of the screen-negative/healthy-control class), defined as
\[
\text{Sensitivity} = \frac{TP}{TP+FN}, \qquad \text{Specificity} = \frac{TN}{TN+FP},
\]
where $TP$, $FN$, $TN$, and $FP$ count true positives, false negatives, true negatives, and false positives among test-set speaker predictions (positive = E-DAIC screen-positive or MODMA MDD). We also report a clinical utility score
\begin{equation}
U = \frac{2\cdot\text{Sensitivity} + \text{Specificity}}{3},
\label{eq:utility}
\end{equation}
where $U \in [0,1]$ (higher is better) weights sensitivity twice as heavily as specificity to represent the chosen asymmetric cost of a missed positive screening case versus a false alarm. This 2:1 weighting is a design choice, not a validated clinical standard, so sensitivity and specificity are also reported separately. A \emph{collapsed} run has (sensitivity, specificity) $=(1,0)$ or $(0,1)$, meaning that it emits the same label for every test speaker. Because both test sets are small, Tables~\ref{tab:results} and~\ref{tab:bybackbone} report F1 as mean$\pm$standard deviation across the six backbones and six architectures, respectively. For the focal subset in Section~\ref{sec:subset}, we additionally compute speaker-level bootstrap 95\% intervals for accuracy and F1 following~\cite{ferrer2022ci}, together with Clopper--Pearson exact intervals for accuracy; all per-seed values are reported in Appendix Table~\ref{tab:ci}.

\begin{table}[!t]
\caption{Mean Metrics by Backbone, Averaged over 6 Architectures ($n=6$ per row; Coll.\ = number of collapsed configurations out of 6 architecture pairings)}
\vspace{-4mm}
\label{tab:bybackbone}
\begin{center}
\scriptsize
\setlength{\tabcolsep}{2pt}
\begin{tabular}{llccccccc}
\toprule
\textbf{Set} & \textbf{Backbone} & \textbf{Acc} & \textbf{F1} & \textbf{AUC} & \textbf{Sens} & \textbf{Spec} & \textbf{$U$} & \textbf{Coll.} \\
\midrule
\multirow{6}{*}{\rotatebox{90}{E-DAIC}}
 & WavLM-Base-Plus    & \textbf{0.775} & \textbf{0.601}$\pm$0.067 & 0.676 & 0.611 & \textbf{0.833} & 0.685 & 0/6 \\
 & WavLM-Large        & 0.710 & 0.465$\pm$0.211 & 0.673 & 0.583 & 0.755 & 0.641 & 1/6 \\
 & HuBERT-Large       & 0.522 & 0.403$\pm$0.192 & 0.582 & 0.722 & 0.451 & 0.632 & 2/6 \\
 & Wav2Vec2-Robust    & 0.268 & 0.416$\pm$0.006 & 0.543 & \textbf{1.000} & 0.010 & 0.670 & \textbf{5/6} \\
 & Data2Vec-Audio-L   & 0.522 & 0.451$\pm$0.031 & 0.611 & 0.722 & 0.451 & 0.632 & 2/6 \\
 & XLS-R-1B           & 0.500 & 0.484$\pm$0.053 & \textbf{0.626} & 0.861 & 0.373 & \textbf{0.698} & 2/6 \\
\midrule
\multirow{6}{*}{\rotatebox{90}{MODMA}}
 & WavLM-Base-Plus    & 0.767 & 0.759$\pm$0.065 & \textbf{0.687} & 0.700 & \textbf{0.833} & 0.744 & 1/6 \\
 & WavLM-Large        & 0.717 & 0.740$\pm$0.084 & 0.640 & 0.767 & 0.667 & 0.733 & 1/6 \\
 & HuBERT-Large       & \textbf{0.800} & \textbf{0.806}$\pm$0.088 & 0.700 & 0.767 & 0.833 & \textbf{0.789} & 1/6 \\
 & Wav2Vec2-Robust    & 0.517 & 0.564$\pm$0.253 & 0.553 & 0.833 & 0.200 & 0.622 & \textbf{5/6} \\
 & Data2Vec-Audio-L   & 0.600 & 0.696$\pm$0.098 & 0.507 & \textbf{0.900} & 0.300 & 0.700 & 3/6 \\
 & XLS-R-1B           & 0.733 & 0.731$\pm$0.049 & 0.600 & 0.700 & 0.767 & 0.722 & 1/6 \\
\bottomrule
\end{tabular}
\end{center}
{\footnotesize Columns as in Table~\ref{tab:results}; each row is the mean over the 6 architectures of Section~\ref{sec:aggregation} for a fixed backbone, with F1 as mean$\pm$standard deviation across those 6 architectures. Coll.\ column matches the per-backbone bars of Fig.~\ref{fig:collapse_backbone}.}
\end{table}

\section{Experimental Results}
\label{sec:results}
Fig.~\ref{fig:pipeline} summarizes the pipeline. We report aggregated performance by architecture and by backbone, the best single configuration per corpus, the collapse phenomenon, and seed sensitivity on a focal subset.
\subsection{Aggregated Performance by Architecture and Backbone}
Table~\ref{tab:results} averages each architecture's metrics across all six backbones, per corpus. On E-DAIC no architecture wins outright: Mean Pooling has the highest accuracy and F1, whereas Statistical Pooling and NetVLAD reach higher sensitivity and utility; NetVLAD's low specificity (0.245) indicates frequent prediction of the positive class. On MODMA, Bi-GRU with Attention has the highest accuracy, F1, ROC-AUC, and utility; the Transformer has the highest sensitivity, and NetVLAD has the highest specificity.
The standard deviations make the robustness point directly: the Transformer encoder's E-DAIC mean of 0.290 carries a spread of 0.206 across backbones and NetVLAD's MODMA mean of 0.632 a spread of 0.302, against 0.042 for Statistical Pooling. The same mean can describe a method that works everywhere or one that alternates between working and failing.

Table~\ref{tab:bybackbone} takes the complementary view: metrics per SSL backbone, averaged over all six architectures. Denoting a metric $M(a,b)$ for architecture $a$ on backbone $b$, Table~\ref{tab:results} reports $\frac{1}{6}\sum_b M(a,b)$ (average over the 6 backbones $b$, for fixed architecture $a$) and Table~\ref{tab:bybackbone} reports $\frac{1}{6}\sum_a M(a,b)$ (average over the 6 architectures $a$, for fixed backbone $b$). The two views agree on the headline conclusion: WavLM-Base-Plus is the strongest, most stable E-DAIC backbone (mean F1 0.601, 0/6 collapsed), while HuBERT-Large edges it out on MODMA (mean F1 0.806 vs.\ 0.759). Wav2Vec2-Robust is the weakest and least stable backbone on both corpora, collapsing on 5 of its 6 architecture pairings per corpus: the clearest single signal that backbone choice can dominate architecture choice.

\subsection{Best Single Configurations}
On the E-DAIC, the best configuration by F1 is Self-Attention pooling on WavLM-Base-Plus (accuracy 0.870, F1 0.667, sensitivity 0.500, specificity 1.000).
On the MODMA, it is Bi-GRU with Attention on Data2Vec-Audio-Large (accuracy 0.900, F1 0.909, sensitivity 1.000, specificity 0.800), a stronger result across all axes, and the architecture is already flagged as most consistent in Table~\ref{tab:results}.

\subsection{The Collapse Phenomenon}
\label{sec:collapse}
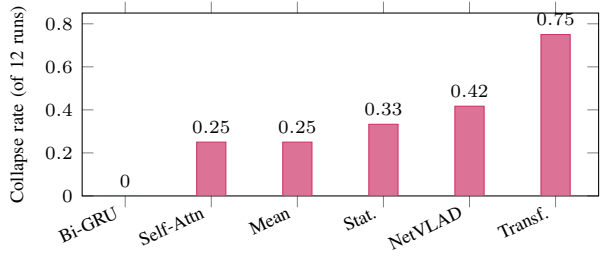
\begin{figure}[!t]
\centering
\begin{tikzpicture}
\begin{axis}[
  width=0.95\columnwidth,
  height=4.0cm,
  ybar,
  bar width=11pt,
  ymin=0, ymax=0.85,
  ylabel={Collapse rate (of 12 runs)},
  symbolic x coords={Bi-GRU,Self-Attn,Mean,Stat.,NetVLAD,Transf.},
  xtick=data,
  x tick label style={font=\scriptsize, rotate=25, anchor=east},
  tick label style={font=\scriptsize},
  label style={font=\scriptsize},
  nodes near coords,
  nodes near coords style={font=\scriptsize},
  every node near coord/.append style={/pgf/number format/.cd, fixed, precision=2},
]
\addplot[fill=purple!55, draw=purple!80] coordinates {
 (Bi-GRU,0.000) (Self-Attn,0.250) (Mean,0.250) (Stat.,0.333) (NetVLAD,0.417) (Transf.,0.750)
};
\end{axis}
\end{tikzpicture}
\caption{Share of runs (of 12: 6 backbones $\times$ 2 corpora) in which an architecture collapses to a single class, under single-seed training. Bi-GRU with Attention never collapses here, but Section~\ref{sec:subset} shows this does not survive seed replication.}
\label{fig:collapse}
\end{figure}

Of the 72 configurations, \textbf{24 (33\%)} collapse to a single predicted class for every test speaker, unevenly spread across architectures (Fig.~\ref{fig:collapse}): the Transformer encoder collapses on 9/12 runs (75\%), NetVLAD on 5/12 (42\%), while Mean Pooling and Self-Attention each collapse on 3/12 (25\%) and Statistical Pooling on 4/12 (33\%). Bi-GRU with Attention is the only architecture that never collapses across these single-seed runs (0/12); Section~\ref{sec:subset} shows this apparent stability is itself seed-dependent.

Collapse is at least as strongly tied to backbone as to architecture. Wav2Vec2-Robust collapses on 10/12 runs (83\%) regardless of pooling head, Data2Vec-Audio-Large on 5/12 (42\%), and WavLM-Base-Plus is the most stable at 1/12 (8\%), as summarized in Fig.~\ref{fig:collapse_backbone}. A benchmark testing only Wav2Vec2-Robust would conclude aggregation for depression detection barely works; one testing only WavLM-Base-Plus would miss the collapse phenomenon almost entirely.

\begin{figure}[!ht]
\centering
\begin{tikzpicture}
\begin{axis}[
  width=0.95\columnwidth,
  height=4.0cm,
  ybar,
  bar width=9pt,
  ymin=0, ymax=0.95,
  ylabel={Collapse rate (of 12 runs)},
  symbolic x coords={WavLM-B+,WavLM-L,HuBERT-L,W2V2-Rob,D2V-L,XLS-R},
  xtick=data,
  x tick label style={font=\scriptsize, rotate=25, anchor=east},
  tick label style={font=\scriptsize},
  label style={font=\scriptsize},
  nodes near coords,
  nodes near coords style={font=\scriptsize},
  every node near coord/.append style={/pgf/number format/.cd, fixed, precision=2},
]
\addplot[fill=teal!55, draw=teal!80] coordinates {
 (WavLM-B+,0.083) (WavLM-L,0.167) (HuBERT-L,0.250) (W2V2-Rob,0.833) (D2V-L,0.417) (XLS-R,0.250)
};
\end{axis}
\end{tikzpicture}
\caption{Share of runs (of 12: 6 architectures $\times$ 2 corpora) in which a given backbone collapses to a single class, computed from the \textbf{Coll.} column of Table~\ref{tab:bybackbone}. Backbone identity predicts collapse at least as strongly as aggregation architecture (Fig.~\ref{fig:collapse}).}
\label{fig:collapse_backbone}
\end{figure}
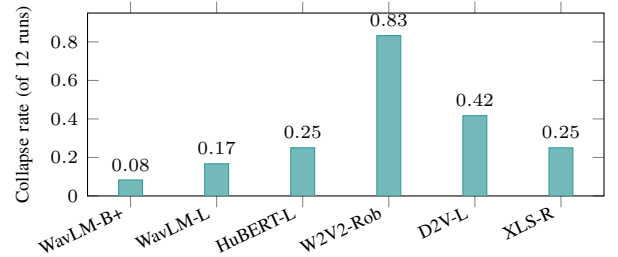 Both conclusions would be artifacts of the one backbone chosen. Reading Table~\ref{tab:results} and Table~\ref{tab:bybackbone} together, neither factor alone predicts outcome: Bi-GRU with Attention stays in a comparatively high, low-variance band across backbones, while the Transformer encoder swings from competitive scores to collapse; it is the \emph{combination} of architecture and backbone that determines whether a configuration works (full per-cell grid released with the code).

\subsection{Seed Sensitivity: A Focal Subset}
\label{sec:subset}
The single-seed grid flags Bi-GRU with Attention as the only architecture that never collapses, but that stability may itself be seed-dependent. We re-ran Bi-GRU and the Transformer encoder on WavLM-Base-Plus and Wav2Vec2-Robust with seeds $\{13,42,87\}$ on both corpora (Table~\ref{tab:subset}), contrasting the factors that looked most and least stable in the grid. On MODMA with WavLM-Base-Plus, Bi-GRU averages F1~$0.11\pm0.16$ and sensitivity 0.07, collapsing to the healthy class on two of three seeds. No focal configuration is both strong and stable, so the single-seed grid understates run-to-run uncertainty.
The intervals in Appendix Table~\ref{tab:ci} overlap across all focal configurations, spanning roughly 0.35 accuracy on E-DAIC and 0.60 on MODMA; at $n=23$ and $n=10$ test speakers no pairwise difference is distinguishable. This reinforces the robustness argument rather than weakening it: single-run point estimates carry little information at these sample sizes. Seeds were fixed in advance for all focal cells; two exploratory runs (seeds 100 and 123), included in the released predictions, give 3/5 collapsed runs at F1~$0.18\pm0.23$, consistent with the subset.

\section{Discussion and Analyses}
\subsection{Robustness vs.\ average performance}
A benchmark reporting only mean F1 per architecture (Table~\ref{tab:results}) would hide that a third of the underlying runs never learn to discriminate at all. Two architectures can have similar means for different reasons: one by reliably landing in a moderate range, another by alternating between strong runs and total collapse. Under single-seed training, only Bi-GRU with Attention never fails outright, but the replication in Section~\ref{sec:subset} shows even that reliability is fragile: no head can be recommended as a default until robustness is established across both backbones and seeds.

\subsection{Why do some architectures collapse more?}
Training uses 75 participants on E-DAIC and 31 on MODMA, so all learned heads operate in a small-sample regime. Collapse is most frequent for the Transformer and NetVLAD in the single-seed grid, but the present experiments do not identify a mechanism. Parameter count alone is insufficient: Bi-GRU is the largest head (approximately 626.0K parameters, excluding the layer featurizer) yet has no collapsed single-seed runs. Plausible contributors include optimization dynamics, class weighting, sample size, and interactions between a head and its frozen embeddings. Distinguishing these explanations requires direct diagnostics and broader multi-seed experiments.

\begin{table}[!t]
\caption{Seed replication on the focal subset (mean$\pm$std over seeds $\{13,42,87\}$). Bi-GRU versus Transformer and WavLM-B+ versus W2V2-Rob contrast architecture and backbone factors that appeared most and least stable in the single-seed grid.}
\vspace{-3mm}
\label{tab:subset}
\begin{center}\scriptsize\setlength{\tabcolsep}{3pt}
\begin{tabular}{lllccc}
\toprule
\textbf{Corpus} & \textbf{Backbone} & \textbf{Arch.} & \textbf{F1} & \textbf{Sens} & \textbf{Spec} \\
\midrule
E-DAIC & WavLM-B+ & Bi-GRU  & 0.405$\pm$0.047 & 0.667 & 0.412 \\
E-DAIC & WavLM-B+ & Transf. & 0.419$\pm$0.007 & 0.833 & 0.235 \\
E-DAIC & W2V2-Rob & Bi-GRU  & 0.375$\pm$0.000 & 0.500 & 0.588 \\
E-DAIC & W2V2-Rob & Transf. & 0.403$\pm$0.015 & 0.889 & 0.118 \\
MODMA  & WavLM-B+ & Bi-GRU  & 0.111$\pm$0.157 & 0.067 & 1.000 \\
MODMA  & WavLM-B+ & Transf. & 0.111$\pm$0.157 & 0.067 & 1.000 \\
MODMA  & W2V2-Rob & Bi-GRU  & 0.667$\pm$0.000 & 1.000 & 0.000 \\
MODMA  & W2V2-Rob & Transf. & 0.667$\pm$0.000 & 1.000 & 0.000 \\
\bottomrule
\end{tabular}
\end{center}
{\footnotesize F1/Sens/Spec defined as in Section~\ref{sec:training}; Sens/Spec columns show the mean over 3 seeds (per-seed values and 95\% bootstrap and Clopper--Pearson intervals in Appendix Table~\ref{tab:ci}).}
\vspace{-6mm}
\end{table}

\subsection{Backbone choice is not a side detail}
Wav2Vec2-Robust collapses in 5 of 6 architecture pairings on each corpus, the highest backbone-specific rate in the grid. This association should not be interpreted as an effect of one pretraining objective: the checkpoints summarized in Appendix Table~\ref{tab:backbones} differ in architecture, scale, training data, pretraining objective, and ASR fine-tuning~\cite{chen2022wavlm,hsu2021robust}. The result instead shows that conclusions about an aggregation architecture depend materially on the backbone with which it is evaluated.

\subsection{MODMA vs.\ E-DAIC}
The highest observed F1 in this grid is higher on MODMA than on E-DAIC (0.909 vs.\ 0.667). The corpora differ in label definition, elicitation protocol, language, population, class balance, and sample size, so this experiment cannot attribute that gap to any single factor. MODMA's test set contains only 10 speakers, meaning that one additional error changes accuracy by 10 percentage points; its result should therefore not be read as evidence that the problem is solved.

\section{Limitations and Ethical Considerations}\label{sec:limitations}
The 72-cell grid uses a single seed, so per-cell metrics cannot separate model effects from run-to-run variation. In the focal three-seed subset, F1 standard deviation reaches 0.157, and the apparent single-seed stability of Bi-GRU does not hold. Small test sets (23 E-DAIC and 10 MODMA speakers) make metrics sensitive to individual participants. For the single-seed grid, validation speakers were merged into training and the reported epoch is the one with the highest test F1, so grid metrics are optimistic upper bounds; the focal multi-seed runs use validation-only checkpoint selection. Mean-pooling within each SSL layer discards frame-level temporal structure, and at the interview level only Bi-GRU encodes clip order: Mean, Statistical, Self-Attention, and NetVLAD are permutation-invariant, and our Transformer has no positional encoding. Because all aggregation heads share one training protocol rather than per-architecture tuning, some failures may reflect interactions with that shared setting. The mechanisms behind collapse, and any broader ranking of backbones, remain open for a full multi-seed study.

Both corpora are de-identified and used under existing agreements. DEPOOL is a research benchmark, not a diagnostic tool: E-DAIC labels are binarized PHQ-8 screening scores, and MODMA's clinician labels do not license deployment. Sample sizes and demographic composition do not support fairness claims, so subgroup evaluation is a priority for future work. Real-world use would require prospective clinical validation, informed consent, and human oversight.

\section{Conclusions}
\label{sec:conclusion}
We presented DEPOOL, pairing six temporal aggregation architectures with six frozen SSL backbones on two depression corpora, yielding 72 configurations with layer selection handled by a learned softmax. Aggregation architecture cannot be evaluated in isolation from backbone or seed: one third of the configurations collapse to single-class prediction, concentrated in the Transformer, NetVLAD, and Wav2Vec2-Robust pairings rather than spread evenly. Bi-GRU with Attention is the only head that never collapses in the single-seed grid, but the focal seed replication breaks that pattern, so we do not recommend a default architecture. The highest-F1 configuration, Bi-GRU on Data2Vec-Audio-Large for MODMA, reaches 0.909, whereas the highest observed E-DAIC F1 is 0.667; performance is corpus-, backbone-, and seed-dependent. Future work should extend multi-seed replication to all 72 cells and evaluate additional backbones, languages, and within-clip temporal encodings.

\begin{figure}[!t]
\centering
\begin{tikzpicture}
\begin{axis}[
  width=0.98\columnwidth,
  height=4.3cm,
  ybar,
  bar width=8pt,
  ymin=-0.05, ymax=1.15,
  ylabel={F1 (mean $\pm$ std over 3 seeds)},
  symbolic x coords={ED-B+-G,ED-B+-T,ED-W2-G,ED-W2-T,MD-B+-G,MD-B+-T,MD-W2-G,MD-W2-T},
  xtick=data,
  x tick label style={font=\tiny, rotate=45, anchor=east},
  tick label style={font=\scriptsize},
  label style={font=\scriptsize},
  error bars/y dir=both,
  error bars/y explicit,
]
\addplot[fill=orange!55, draw=orange!85] coordinates {
 (ED-B+-G,0.405) +- (0,0.047)
 (ED-B+-T,0.419) +- (0,0.007)
 (ED-W2-G,0.375) +- (0,0.000)
 (ED-W2-T,0.403) +- (0,0.015)
 (MD-B+-G,0.111) +- (0,0.157)
 (MD-B+-T,0.111) +- (0,0.157)
 (MD-W2-G,0.667) +- (0,0.000)
 (MD-W2-T,0.667) +- (0,0.000)
};
\end{axis}
\end{tikzpicture}
\vspace{-3mm}
\caption{Seed replication on the focal subset of Table~\ref{tab:subset} (ED = E-DAIC, MD = MODMA, B+ = WavLM-Base-Plus, W2 = Wav2Vec2-Robust, G = Bi-GRU with Attention, T = Transformer encoder). Error bars are one standard deviation over 3 seeds. The widest bars, both on MODMA with WavLM-Base-Plus, show that single-seed stability does not imply multi-seed stability.}
\label{fig:seed_variance}
\vspace{-4mm}
\end{figure}
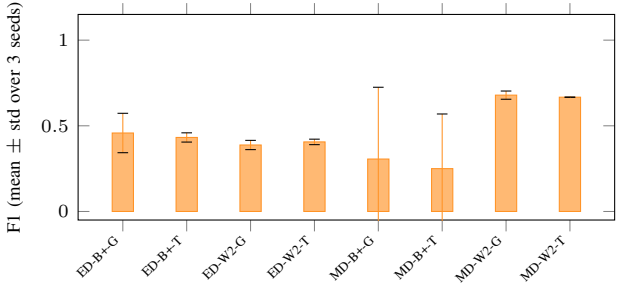

\section*{Acknowledgment}
This research was funded, in part, by the Advanced Research Projects Agency for Health (ARPA-H). The views and conclusions contained in this document are those of the authors and should not be interpreted as representing the official policies, either expressed or implied, of the U.S. Government.
This work used Jetstream2 through ACCESS allocation CIS260814 and CIS260813 also was supported by the National Science and Technology Council, Taiwan, under Grant 114-2917-I-564-030 (Huang-Cheng Chou).
OpenAI Codex and Claude Fable were used to assist with language editing and camera-ready compliance checks; all scientific claims, experimental data, analyses, and conclusions remain the authors' responsibility.

\bibliographystyle{IEEEtran}
\bibliography{IEEEexample}

\clearpage
\appendix

\section{Supplementary Material}

Table~\ref{tab:backbones} summarizes the six self-supervised learning (SSL) speech backbones evaluated within the DEPOOL benchmark. We select these models to represent diverse pretraining objectives, including contrastive learning, masked prediction, and self-distillation, as well as varying parameter counts and training data sizes.
The backbones range from the 94M-parameter WavLM-Base-Plus to the 965M-parameter XLS-R-1B, encompassing both purely self-supervised models and those further fine-tuned for automatic speech recognition (ASR-FT).
For each backbone, we provide the number of available Transformer hidden-state layers, $L$, and the hidden dimension, $D$, both of which are used by the semi-fine-tuned layer-aggregation protocol defined in Section~\ref{sec:featurizer}. All backbones are kept frozen throughout our training process, ensuring that the benchmark results reflect the downstream aggregation architecture and the learned layer-weighting efficacy rather than backbone-internal weight updates.

\begin{table*}[!t]
\caption{SSL Backbones Evaluated in DEPOOL.}
\vspace{-4mm}
\label{tab:backbones}
\begin{center}
\begin{tabular}{lccccc}
\toprule
\textbf{Backbone} & \textbf{Layers} & \textbf{Dim.} & \textbf{Params} & \textbf{Pretrain.} & \textbf{Ref.} \\
\midrule
WavLM-Base-Plus      & 13 & 768  & 94M  & SSL, 94k~h Libri-Light/GigaSpeech/VoxPopuli & \cite{chen2022wavlm} \\
WavLM-Large          & 25 & 1024 & 316M & SSL, 94k~h Libri-Light/GigaSpeech/VoxPopuli & \cite{chen2022wavlm} \\
HuBERT-Large         & 25 & 1024 & 316M & SSL on 60k~h Libri-Light, then ASR-FT on 960~h LS & \cite{hsu2021hubert} \\
Wav2Vec2-Robust      & 25 & 1024 & 317M & SSL, domain-mixed (read + noisy + telephone speech) & \cite{hsu2021robust} \\
Data2Vec-Audio-Large & 25 & 1024 & 314M & SSL on 60k~h Libri-Light, then ASR-FT on 960~h LS & \cite{baevski2022data2vec} \\
XLS-R-1B             & 49 & 1280 & 965M & SSL, 436k~h multilingual (128 languages) & \cite{babu2021xlsr} \\
\bottomrule
\end{tabular}
\end{center}
\vspace{-2pt}
{\footnotesize SSL = self-supervised only; SSL+FT = self-supervised then ASR fine-tuned (HuBERT-Large: \texttt{hubert-large-ls960-ft}; Data2Vec-Audio-Large: \texttt{data2vec-audio-large-960h}). Layers = number of Transformer hidden-state outputs, including the feature-extractor output, available to the featurizer of Eq.~\eqref{eq:featurizer}. Dim.\ = hidden size $D$ of each layer. Params = approximate total parameter count of the released checkpoint (frozen; not updated during our training). LS = LibriSpeech.}
\end{table*}

Table~\ref{tab:ci} reports the per-seed confidence intervals for the focal subset of Table~\ref{tab:subset}. Speaker-level 95\% bootstrap intervals for accuracy and positive-class F1 are computed with the toolkit of~\cite{ferrer2022ci}; Clopper--Pearson exact intervals are reported for accuracy.

\begin{table*}[!t]
\caption{Per-seed 95\% confidence intervals for the focal subset of Table~\ref{tab:subset}. Boot.\ = speaker-level bootstrap; CP = Clopper--Pearson exact (accuracy only).}
\vspace{-4mm}
\label{tab:ci}
\begin{center}
\scriptsize
\setlength{\tabcolsep}{4pt}
\begin{tabular}{lllccccc}
\toprule
\textbf{Corpus} & \textbf{Backbone} & \textbf{Head} & \textbf{Seed} & \textbf{Acc} & \textbf{Acc 95\% Boot.} & \textbf{Acc 95\% CP} & \textbf{F1 (95\% Boot.)} \\
\midrule
\multirow{3}{*}{E-DAIC} & \multirow{3}{*}{WavLM-B+} & \multirow{3}{*}{Bi-GRU} & 13 & 0.609 & [0.435, 0.783] & [0.385, 0.803] & 0.471 [0.143, 0.737] \\
 & &  & 42 & 0.391 & [0.174, 0.609] & [0.197, 0.615] & 0.364 [0.105, 0.615] \\
 & &  & 87 & 0.435 & [0.217, 0.652] & [0.232, 0.655] & 0.381 [0.100, 0.640] \\
\midrule
\multirow{3}{*}{E-DAIC} & \multirow{3}{*}{WavLM-B+} & \multirow{3}{*}{Transf.} & 13 & 0.652 & [0.477, 0.826] & [0.427, 0.836] & 0.429 [0.000, 0.714] \\
 & &  & 42 & 0.261 & [0.087, 0.435] & [0.102, 0.484] & 0.414 [0.160, 0.606] \\
 & &  & 87 & 0.261 & [0.087, 0.436] & [0.102, 0.484] & 0.414 [0.160, 0.607] \\
\midrule
\multirow{3}{*}{E-DAIC} & \multirow{3}{*}{W2V2-Rob} & \multirow{3}{*}{Bi-GRU} & 13 & 0.565 & [0.390, 0.783] & [0.345, 0.768] & 0.375 [0.000, 0.641] \\
 & &  & 42 & 0.565 & [0.348, 0.739] & [0.345, 0.768] & 0.375 [0.000, 0.637] \\
 & &  & 87 & 0.565 & [0.348, 0.783] & [0.345, 0.768] & 0.375 [0.000, 0.667] \\
\midrule
\multirow{3}{*}{E-DAIC} & \multirow{3}{*}{W2V2-Rob} & \multirow{3}{*}{Transf.} & 13 & 0.435 & [0.261, 0.652] & [0.232, 0.655] & 0.381 [0.111, 0.640] \\
 & &  & 42 & 0.261 & [0.087, 0.435] & [0.102, 0.484] & 0.414 [0.160, 0.606] \\
 & &  & 87 & 0.261 & [0.087, 0.435] & [0.102, 0.484] & 0.414 [0.160, 0.606] \\
\midrule
\multirow{3}{*}{MODMA} & \multirow{3}{*}{WavLM-B+} & \multirow{3}{*}{Bi-GRU} & 13 & 0.600 & [0.300, 0.900] & [0.262, 0.878] & 0.333 [0.000, 0.770] \\
 & &  & 42 & 0.500 & [0.200, 0.800] & [0.187, 0.813] & 0.000 [0.000, 0.000]$^\dagger$ \\
 & &  & 87 & 0.500 & [0.200, 0.800] & [0.187, 0.813] & 0.000 [0.000, 0.000]$^\dagger$ \\
\midrule
\multirow{3}{*}{MODMA} & \multirow{3}{*}{WavLM-B+} & \multirow{3}{*}{Transf.} & 13 & 0.600 & [0.300, 0.900] & [0.262, 0.878] & 0.333 [0.000, 0.800] \\
 & &  & 42 & 0.500 & [0.200, 0.800] & [0.187, 0.813] & 0.000 [0.000, 0.000]$^\dagger$ \\
 & &  & 87 & 0.500 & [0.200, 0.800] & [0.187, 0.813] & 0.000 [0.000, 0.000]$^\dagger$ \\
\midrule
\multirow{3}{*}{MODMA} & \multirow{3}{*}{W2V2-Rob} & \multirow{3}{*}{Bi-GRU} & 13 & 0.500 & [0.200, 0.800] & [0.187, 0.813] & 0.667 [0.333, 0.889] \\
 & &  & 42 & 0.500 & [0.200, 0.800] & [0.187, 0.813] & 0.667 [0.333, 0.889] \\
 & &  & 87 & 0.500 & [0.200, 0.800] & [0.187, 0.813] & 0.667 [0.333, 0.889] \\
\midrule
\multirow{3}{*}{MODMA} & \multirow{3}{*}{W2V2-Rob} & \multirow{3}{*}{Transf.} & 13 & 0.500 & [0.200, 0.800] & [0.187, 0.813] & 0.667 [0.333, 0.889] \\
 & &  & 42 & 0.500 & [0.200, 0.800] & [0.187, 0.813] & 0.667 [0.333, 0.889] \\
 & &  & 87 & 0.500 & [0.200, 0.800] & [0.187, 0.813] & 0.667 [0.333, 0.889] \\
\bottomrule
\end{tabular}
\end{center}
\vspace{-2pt}
{\footnotesize Backbone and head abbreviations as in Fig.~\ref{fig:seed_variance}. $^\dagger$ Collapsed run predicting the healthy class for every speaker, so positive-class F1 is exactly zero in every resample and its bootstrap interval degenerates to a point. This reflects the absence of resampling variability, not genuine certainty about the underlying F1; the accompanying accuracy intervals for these rows remain wide. Identical accuracy counts give identical Clopper-Pearson intervals by construction; bootstrap intervals for the same point estimate can differ slightly because resampling is stochastic.}
\end{table*}

\end{document}